# Microfabrication of a 16 MHz 1D-pMUT-Array for photoacoustic and ultrasound imaging

Atheeth. S[1], Chandrashekar L N[2], Isha Munjal[3], Swathi Padmanabhan[3], Jaya Prakash[3], Manish Arora[1]

[1]Department of Design and Manufacturing (DM), Indian Institute of Science (IISc)
[2]Centre for Nano Science and Engineering (CeNSE), Indian Institute of Science (IISc)
[3]Instrumentation and Applied Physics (IAP), Indian Institute of Science (IISc)

*Abstract*—In this paper, we describe the complete fabrication process of a 1D piezoelectric Micromachined Ultrasound Transducer (pMUT) array operating at 16 MHz underwater. We demonstrate the applicability of this pMUT Array in medical imaging using photoacoustic imaging (PAI) and ultrasound imaging (USI) experiments. There are 16 individual pMUT devices in the array, the radius of each device is 25 microns with a pitch of 100 microns (centre-to-centre). A 1-micron thick AlN (Aluminium Nitride) thin film is the piezoelectric material of choice for our pMUT array. This thin film was achieved by improving upon the control parameters in RF magnetron sputtering process. The working of this pMUT was validated by performing optical, electrical, and acoustic characterization. The 1D pMUT array was characterized optically using Laser Doppler Vibrometer (LDV) wherein the pMUT membrane showcased displacement of 6.2 pm/V for the in-air measurements at resonance of 20 MHz, the resonance frequency underwater was 16.2 MHz, with an effective displacement of 2.08 pm/V. Electrical characteristics were obtained through lock-in amplifier measurements, these were in close match to LDV results. Acoustical characteristics of the array was obtained through imaging experiments.

## I. INTRODUCTION

Miniature ultrasound (US) imaging probes for Intra-Cardiac Echocardiography (ICE) and Intra Vascular ultrasound (IVUS) are currently in clinical practice. These probes are derived from conventional ultrasound material and are complex to manufacture. Applications such as minimally invasive surgeries require probes to enable viewing through tiny surgical openings. Besides, these procedures invariably need to employ high frequency US to visualize fine (small) structures such as nerves and blood vessels around a tumour site via B-mode images acquired by pulse-echo techniques. Specialized transducers must be designed for such applications [1].

Micromachined US Transducer (MUT) has the potential to develop a probe with small form-factor (<2 mm diameter) capable of imaging at high-frequency (>15 MHz) [1]. Based on the actuation scheme, MUT's are classified as cMUT (capacitive) and pMUT (piezoelectric) [1], each with its own set of advantages and disadvantages. pMUT's could potentially be the appropriate technology for high frequency array probes suitable for minimally invasive surgeries. Previous review articles on pMUT have provided an extensive survey of pMUT applications, governing equations for pMUT design, material selection and fabrication processes, design limitations and challenges with pMUT-array development [1].

The motivation for considering pMUT for high-frequency imaging is the lower operating voltages of the devices and the design freedom. Structural geometry and material properties can be varied as per need. cMUT has already been commercially proven as a medical imaging technology, albeit its high operating voltage currently limits its use at higher frequencies. Therefore, we have chosen to focus on pMUT's for our device development.

The pMUT fabricated through MEMS processes are then a potential alternative to the conventional ultrasound transducers for the ultrasound imaging application. In literature, pMUT has been demonstrated for many applications such as gesture recognition, range detection, fingerprint sensing and imaging among others [2]–[6]. In our work we focus on the multi modal capabilities of pMUT that can be leveraged for both photoacoustic and ultrasound imaging system development. Subsequently we list the advantages of pMUT reported in prior literature and highlight a few research opportunities in pMUT device development. At the core of fabricating a pMUT is the careful consideration for choosing and developing the right piezoelectric material. Higher quality piezoelectric thin films can significantly push the development of biomedical imaging probes [7], [8].

pMUT has the potential to achieve high-volume imaging and a better coupling to the imaging medium without the need for matching layer. Hence using pMUT; high-density, low-cost, lower-power consuming imaging probes are a possibility [9], this is the hypothesis we wish to build upon in the context of multi-modal imaging.

Breakthrough products have been created based on CMUT (Capacitive Micromachined Ultrasound Transducer) technology. Depending on the piezoelectric material used in the pMUT, we can monolithically integrate the pMUT device onto the CMOS (Complementary Metal Oxide Semiconductor) layer as reported [10], [11]. High capacitance of pMUT elements combined with lower electrical impedance leads to better matching with the support electronics and makes the device less sensitive to parasitic. A lower dielectric constant of Aluminium Nitride (AlN) in comparison to Lead Zirconium Titanate (PZT) allows for a superior performance in terms of a receive sensitivity value [12], which was the leverage for our photoacoustic (and ultrasound) imaging experiment. To obtain a high-quality medical imaging capability, transmit sensitivity of the pMUT needs improvement. Less number of studies have reported to improve the transmit efficiency [13]. In the context of fingerprint sensing, pMUT's are known to provide good image quality [14]. Few studies have focussed on beam modelling, pressure field mapping, successful transmission, and reception of ultrasound by 2D pMUT arrays [15].

Over the last 15 years (2010 onwards) Photoacoustic Imaging (PAI) has gained prominence in research. Continuous improvements in image reconstruction algorithms and instrumentation have enabled PAI to be developed and validated for many applications. In the following works, PAI right from lab-level experiments to pre-clinical studies are reported [16]–[22]. Previous seminal works that report on ultrasound and photoacoustic imaging using pMUT can be found in Refs [24]–[35].

Structurally, a pMUT is a multi-layered plate typically with one piezoelectric layer sandwiched between two metal layers and one passive base layer. This structure is also referred to as a unimorph structure. The metal layers act as electrodes for applying voltage to actuate the pMUT. An electric field across the piezoelectric layer creates in-plane stresses along the piezoelectric material. The developed stresses are not symmetric about the neutral plane of the piezoelectric layer. The neutral plane of piezoelectric layer is at a certain distance above the neutral plane of base passive layer. Therefore, a net bending moment is caused around the axis. A sinusoidal input ensures flexural vibrations from the unimorph, this gets coupled to the medium as ultrasonic signal in transmit mode of operation. Conversely, an acoustic signal impinging on the structure induces vibrations, developing a potential difference across its piezoelectric layer.

The best response of the pMUT, as a receiver and transmitter, is obtained when the signal matches its resonant frequency. The pMUT can be directly coupled with the fluid medium (unlike in a conventional transducer). The structural dissipation becomes exceedingly small as compared to acoustic dissipation. This is due to the small size of the device. Highly efficient conversion from electrical energy to acoustic energy happens due to mismatch in dissipation values.

In the subsequent sections we will describe the fabrication process followed to achieve the high frequency pMUT characterization and associated experimental work to demonstrate the applicability of the fabricated pMUT array for multi-modal imaging is presented in later sections.

## II. pMUT FABRICATION PROCESS

Earliest report of pMUT fabrication was in 1997 [36]. Different research groups also presented their fabrication methods subsequently [37]–[42]. More recently pMUT's are fabricated based on Silicon-On-Insulator (SOI) wafers, however this increases the cost of fabrication. Hence in our work, pMUT was fabricated on Silicon (Si) wafer.

The design of pMUT was dictated by the requirements of the application (having small form factor and high frequency). Theoretical calculations were performed to arrive at the desired dimensions for the device, followed by COMSOL Multiphysics simulation model. The simulation model allowed us to analyse the working of our pMUT design in-air and underwater. The results from the simulations were referenced when device characterization was performed in optical, electrical, and acoustic domains, results of which are presented in next section of the paper.

Each fabricated pMUT was having a 25-micron radius. Platinum (Pt) was used as the bottom electrode for all the devices in the array. The top electrode constructed out of Titanium-platinum (Ti-Pt) was made 65% of the bottom electrode dimension, this increased the central displacement values and ensured there were no electrical shorts between the electrodes. Sandwiched between the electrodes is a 1-micron thick AlN thin film grown using RF magnetron sputtering process. Figure 1 depicts the fabrication process followed in this work.

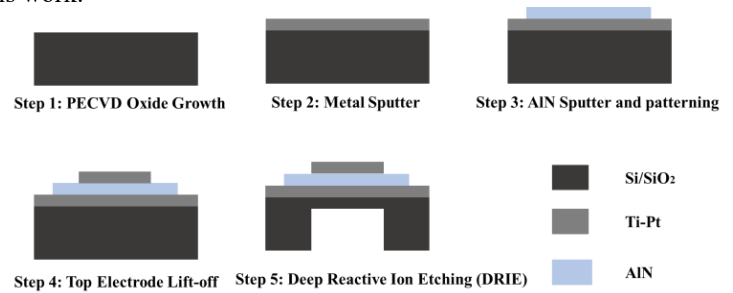

Figure 1: PMUT Fabrication Process Flow

A 3-inch n-type double side polished, <100> oriented Silicon (Si) wafer is the base layer. 1-micron thick $SiO_2$ was deposited on both sides of the wafer through Plasma Enhanced Chemical Vapour Deposition (PECVD) process, this oxide layer eliminated the shorting of bottom electrode with the base layer. 200 nm thick Ti-Pt bottom electrode was patterned using photo lithography process and the bottom-layer mask design, after which the exposed electrode was etched using a diluted aqua regia solution at 70°C.

1-micron AlN thin film was deposited by RF Magnetron Sputtering, the process of which is described in detail by author's previous work [43]. AlN layer was patterned using the second mask, the etchant used was orthophosphoric acid heated to 80°C. Third mask of the process was used to pattern the top electrode, which was designed smaller than the bottom electrode and AlN radii. Lift-off was performed in acetone solution for 2 minutes 40 seconds, this completed the processes on topside of the wafer, the stack achieved is Si/SiO$_2$/Ti-Pt/AlN/Al-Au.

The top layers were coated with AlN paste before removing the backside SiO$_2$, photolithography was performed using the fourth mask to open the etch windows. Through the process of Reactive Ion Etching – Fluorine (RIE-F), SiO$_2$ hard mask was etched. After this etch, base layer Si needs to be etched to form the membrane support layer.

Most crucial step of the entire process was Deep Reactive Ion Etching (DRIE) wherein Si layer was etched with precision. A 50-micron diameter was etched from the backside of the wafer to a depth of 325-micron, leaving behind 2 – 5 microns of base Si as the support for the released membranes. An etch rate of 16µm/min was observed. Stack thickness and the etched diameter determine the resonance frequency of the pMUT which was at 20 MHz in-air and 16 MHz underwater. Thickness of the materials in the stack were controlled through the deposition parameters.

The opening for the through etch here is of 50-micron diameter and we need to etch out 350 micron of Si depth wise, the aspect ratio of the etch being 1/7, is at the limit of DRIE process, hence we modified the process slightly to achieve the desired through-etch. The modification involved changing the flow rate and pressure of etching gases midway through the deep trench etch. Each of the process steps mentioned above went through carefully designed experimentation to achieve repeatable results. The fabrication process runs also highlighted the need for high-quality piezoelectric thin film deposition, which is at the core of high-performance pMUT device development.

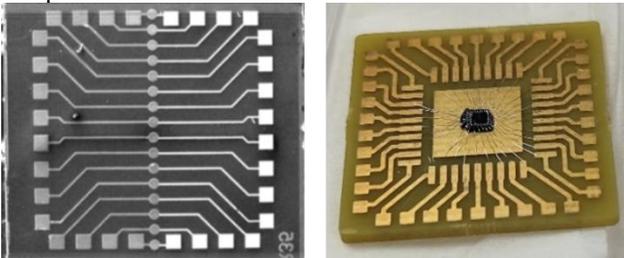

Figure 2: SEM Image of fabricated array and array-die measuring 1.5mm*1.5mm wire-bonded onto a custom-made PCB

Figure 2 indicates the scanning electron microscope (SEM) image showing top view of the fabricated linear pMUT Array. There are 16 individual pMUT devices in the array, the radius of each device is 25 microns with a pitch of 100 microns (centre-to-centre as shown in Figure 2, each element had two metal electrode layers: one for ground connection and other for signal. This array is housed on a chip of size 1.5 mm x 1.5 mm.

The fabricated pMUT devices were die-bonded onto a custom-made PCB using epoxy EPO-TEK-HC-70 as shown in Fig. 3. Wire-bonding was performed from the electrodes onto the die/chip to PCB pads. A parylene layer of 500 nm thickness was deposited on the die, masking tape was used to protect the PCB pads during the parylene deposition step. Wires were soldered on PCB pads, connecting voltage supply, or measuring equipment. Since the pMUT had to be tested for underwater response in imaging experiments, epoxy coating was carried out to ensure that the wires were not exposed to water.

Once the array was fabricated, it was tested for its performance in terms of electrical, optical and acoustic characterizations. More details on the characterization of these devices are presented in the next section of the paper.

### III. pMUT CHARACTERIZATION

a) Laser Doppler Vibrometer (LDV) Measurements

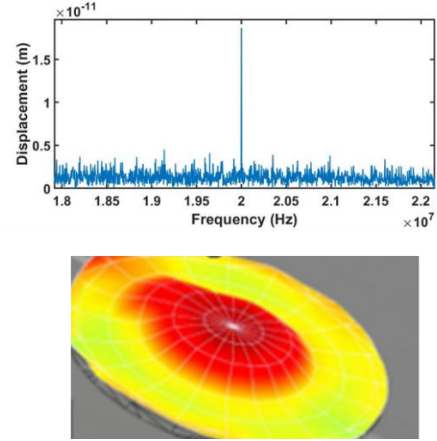

Figure 3: In-air LDV measurement with peak displacement of 18.7pm at 20 MHz and corresponding mode shape

The wire-bonded pMUT array was supplied an operating voltage of 2V AC and probed by a laser incident on the central portion of the pMUT element. In this characterization method, we measured the central displacement values of the pMUT across a range of input frequencies. At the resonance frequency of the device, maximum central displacement is expected. This characterization was performed both in-air and underwater.

In-air LDV measurements is indicated in Fig. 4, wherein a maximum displacement of 18.7 pm at 20 MHz was observed when the pMUT was actuated by 3V input. Fig. 5 shows the mode shape at resonance, as expected the region covered by top-electrode showed the maximum displacement. In the underwater test, a maximum displacement of 2.08 pm/V was observed at resonance frequency of 16.18 MHz.

b) Lock-In Amplifier Measurements:

The principle of lock-in measurements relies on frequency mixing of output signal from the Device Under Test (DUT) and a reference signal from the amplifier. The output in the frequency domain due to the mixing will show a response at resonance frequency of the device plus the reference frequency. Figs 6 and 7 show that the resonance values (at maximum output voltage and phase change) are in close agreement with LDV values. This corroborates well with the COMSOL simulation results indicated in Table-1.

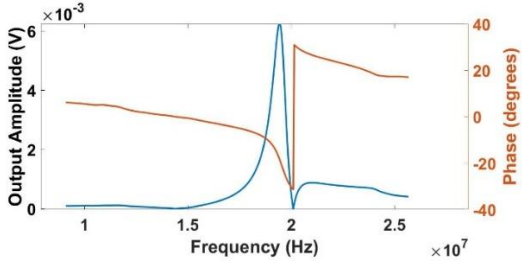

Figure 4: In-Air Lock-In amplifier measurement

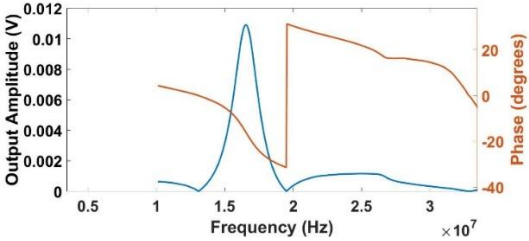

Figure 5: Underwater Lock-in amplifier measurements

Table 1: Resonance Frequency value comparison (experimental v/s simulation)

| Resonance | LDV | Lock-In | Simulation |
|---|---|---|---|
| In-air | 20 | 19.8 | 18 |
| Underwater | 16.2 | 16.6 | 16.5 |

b) Transmit Sensitivity measurement using hydrophone:

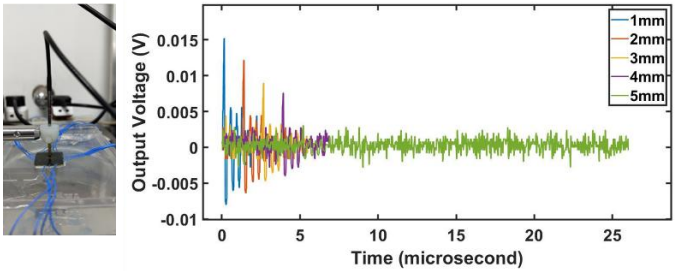

Figure 6: Hydrophone setup and data

Fig. 6 shows the output signal from the Onda HNR 1000 hydrophone when the pMUT was actuated using a function generator with 5V single sinewave burst at a frequency of 16.2 MHz, the entire setup of pMUT and hydrophone was underwater. The hydrophone sensitivity data (398nV/Pa) was used to convert the output voltage from the hydrophone to a corresponding output pressure amplitude. The mapping between the output voltage and output pressure for up to 5mm distance is shown in Table-2.

Table 2: Output Pressure Value comparison (experimental v/s simulation)

| Distance (mm) | COMSOL Simulation (kPa) | Experimental (kPa) | Corresponding Output Voltage (mV) |
|---|---|---|---|
| 1 | 42 | 38 | 15 |
| 2 | 35 | 30 | 12 |
| 3 | 32 | 23 | 9 |
| 4 | 29 | 19 | 7.5 |
| 5 | 26 | 15 | 3 |

## IV. IMAGING EXPERIMENTS

After device characterization was performed and documented, the next set of experiments was to check for the pMUT capabilities for multi-modal imaging. Hence, we did both the photoacoustic and ultrasound imaging experiments. In the photoacoustic experiment, pMUT were evaluated as a receiver of ultrasound signals, whereas in the ultrasound experiment, we note the transmit and receive capabilities of the fabricated pMUT Array.

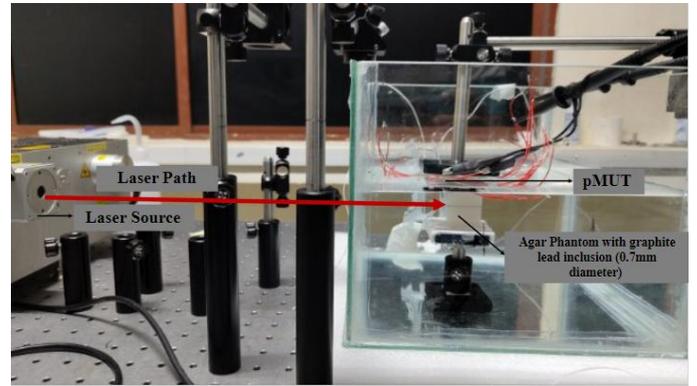

Figure 7: Photoacoustic Imaging Setup

In our photoacoustic experiment, an agar-based phantom with a 0.7 mm diameter graphite lead inclusion was used as the target/object to be imaged. A pulsed laser with the following specifications: INNOLAS SpitLight 1000 OPO-532 with the wavelength range: 660nm-2500nm and pulse width: 7ns along with a repetition rate of 30Hz was used as the source. The laser was set at 740nm (the absorption range for graphite lead material). This laser energy is absorbed by the inclusion, the resulting response of thermo-elastic expansion and contraction results in an emission of a pressure pulse at the the ultrasound frequencies [44]-[49]. The generated photoacoustic signals will be picked up by the fabricated pMUT, which is used as a receiver in this experiment. The receive response was captured by varying distance between pMUT and inclusion, across all the 16 channels on the pMUT array. The photoacoustic experiment setup is shown on Fig. 7.
The reconstructed images are shown in Fig. 8.

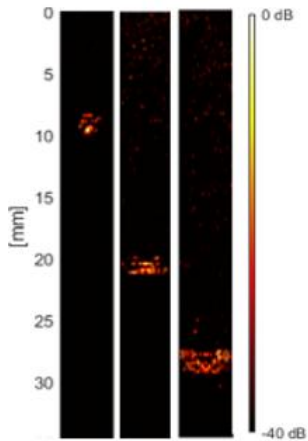

Figure 8: Reconstructed Photoacoustic images at 10, 20, 32 mm respectively (left to right)

To reconstruct photoacoustic images from the received data, we edited the time delay calculations in MATLAB UltraSound Toolbox (MUST) and set transmit time = 0 and implemented the standard delay and sum algorithm on the received data from all channels of the array.

Some of the most recent literature on ultrasound imaging using pMUT are [24], [27], [31], [32]. In comparison to these papers, our work demonstrates the feasibility of high-frequency (16 MHz) ultrasound imaging using the fabricated 1D-pMUT-array. We have used synthetic aperture imaging scheme to capture and reconstruct image frames. In each ultrasound experiment, one element of the pMUT is actuated and receive signals are captured using the rest of the elements of the array. In each experiment, 16 such frames are captured to form one summed frame. The resultant summed frame is represented in Fig. 9 and Fig. 10, respectively.

In experiment 1: a single 2 mm diameter copper wire was included in an agar phantom. In experiment 2: three copper wires of 0.5 mm diameter each were placed in the phantom for imaging. The reconstruction of the received ultrasound data was performed using the delay and sum algorithm. In transmit since one element is active at a time, the rest of the elements were set as OFF. On the receive, the rest 15 elements were active for reception of the reflected echoes.

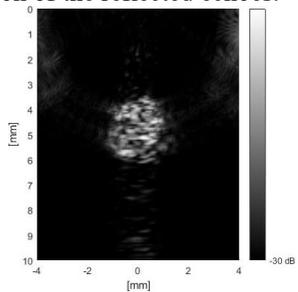

Figure 9: Summed Image Frame with 2mm diameter wire

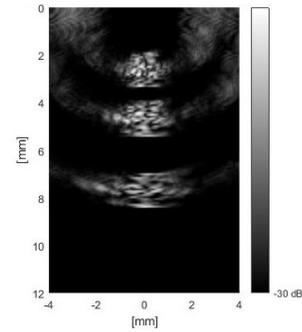

Figure 10: Summed Image frame with three wires of 0.5mm diameter each

The experimental demonstration was also supplemented with simulations. MUST contains two important functions: PFIELD and SIMUS, where the acoustic pressure field generated from the defined transducer elements are simulated in the defined domain, this pressure field will interact with the objects set to scatter, reflect ultrasound and these reflected echoes then reach the transducer elements, SIMUS computes the necessary time delays and helps generate realistic ultrasound images from the received RF data. Note that MUST solve the wave equation in Fourier domain to obtain the RF signals [50].

The imaging simulation environment was modified to mimic the experimental scenario of imaging copper wires of different diameters (2mm and 0.5mm) embedded in an agar-based phantom. Note that the imaging simulation emulates a synthetic aperture imaging scheme (one element of the array transmits ultrasound at a time and the rest 15 elements receive the reflected echoes) for the one wire and three wire scenarios. MUST functions were modified to create 16 different frames for transmit of each individual element.

In the simulations: consider the transmission from one element, at that instant rest of the elements should be OFF. On receive, only 15 elements should be active. These required changes to the transmit and receive delay calculations. Within the gamut of functions provided in MUST, we modified function txdelay and rxdelay to correspondingly change transmit and receive delay calculations.

Consider total travel time of the ultrasound wave from transmit to being received on elements of the array to be $t_{overall}$. Then, $t_{overall}$ = travel time to scatter object ($t_t$) + travel time from object to receive element ($t_r$). To implement Delay and Sum (DAS) algorithm to form image, we calculate the following values: speed of sound in medium (c), distance travelled (d) and $t_{overall}$. Knowing two of them the other can be calculated as $c = d/t_{overall}$. Distance for subsequent elements will be sqrt ($d^2$ + $pitch^2$) with pitch = 100-micron.

B-mode data from all the 16 frames were summed and averaged to form the resultant frames shown in Fig 11. Simulation results for the three-wire scenario present a summed frame as shown below in Fig 11:

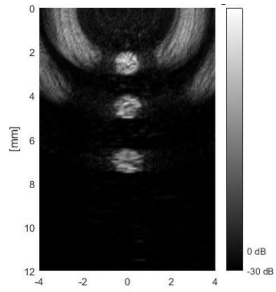

Figure 11: Simulation of the three-wire scenario

## V. Discussion

A 1D pMUT Array was fabricated and subsequently characterized using optical (using SEM), electrical (Locking measurements), and acoustic methods (LDV measurements). Choosing the right material stack was a critical decision during the pMUT fabrication procedure. This pMUT at the core has a highly c-axis oriented AlN piezoelectric thin film, which will enable improved sensitivity to receive weak photoacoustic signals from depth >25mm. Detailed experiments were carried out to obtain this piezoelectric thin film of AlN. RF Magnetron Sputtering of this AlN thin film was reported in a previous work [69]. There is further scope for improving the AlN thin film quality by considering a thorough investigation of all the parameters that affect the thin film growth.

A key test after fabricating the pMUT was to note the response to underwater actuation and record the output through a hydrophone. In this test, we observed a maximum output pressure of 38 kPa at 1 mm distance from the transducer when actuated by 3V. This output pressure was in the range of previously reported values. A 6.2 pm/V in-air displacement of the vibrating pMUT membrane improved upon the previously reported values.

Imaging simulations validated the design suitability of the array for medical imaging applications. However, from Figure 14 we note that there are artifacts present in the simulated image. This work also does not consider the effect on image quality output if other beamforming methods (DMAS, MVDR etc) are used as a post-processing step. Imaging experiments highlight the applicability of using this high-frequency pMUT-array as a multi-modal imaging device (photoacoustic + ultrasound).

Imaging an object of size 0.5. mm at a depth of 6 mm from the transducer demonstrates the usefulness of high-frequency imaging in visualizing smaller structures. Further improvements can be made to check if the array can pick up poor acoustic contrast as opposed to just copper wire set in agar.

Further improvements in piezoelectric thin-film deposition coupled with efforts to reduce residual stresses in the material stack can result in superior imaging performance of the pMUT. Enabling the surgeons to visualize blood vessels and nerves around a tumor allows for preserving vasculature in the brain during surgery is a possibility through system development around the fabricated pMUT.

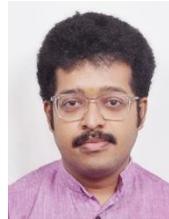

**Atheeth. S:** Atheeth recently presented his Ph D Colloquium on the thesis titled "Design, Simulation, Microfabrication and Characterization of a 16 MHz 1D-pMUT-Array" at Department of Design and Manufacturing (DM), Indian Institute of Science (IISc). He has a bachelor's degree in Electronics and Communication from Dr. Ambedkar Institute of Technology followed by a master's degree in biomedical Signal Processing and Instrumentation from BMS College of Engineering. His research interests include ultrasound imaging, ultrasound transducer development and pMUT's.

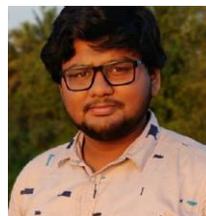

**Chandrashekar L N:** A Doctoral Student at Tyndall National Institute, Cork, Ireland. He holds his bachelor's degree in Electronics and Communication Engineering from Nitte Meenakshi Institute of Technology, Bengaluru. His research focuses on the fabrication and characterization of field-effect transistors for next-generation high-performance devices, with applications in quantum sensing.

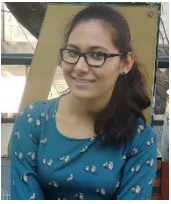
**Isha. M:** Isha Munjal is a PhD student in Department of Instrumentation and Applied Physics at Indian Institute of Science, Bangalore. She received her master's degree in physics from SLIET University. Her current research focuses on deep learning-driven advances in medical imaging modalities, including computed Tomography and photoacoustics tomography.

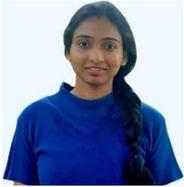
**Swathi. P:** Swathi Padmanabhan is a Ph D student at the Department of Instrumentation and Applied Physics at Indian Institute of Science, Bangalore. She received her bachelor's degree in technology on Applied electronics and Instrumentation from APJ Abdul Kalam Kerala Technological University. Currently, her research focuses on non-invasive polarization enhanced photoacoustic sensing of biomolecules and photo-theranostic approaches such as photothermal therapy.

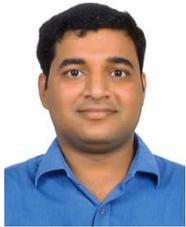
**Jaya Prakash:** Jaya Prakash received the B.Tech. degree in information technology from the Amrita School of Engineering, Bengaluru, India, in 2010, and the M.Sc. degree in engineering and the Ph.D. degree in medical imaging from the Indian Institute of Science, Bengaluru, in 2012 and 2014, respectively. Prior to his current position as an Assistant Professor with the Department of Instrumentation and Applied Physics, Indian Institute of Science, he was an Alexander von Humboldt Fellow at the Institute for Biological and Medical Imaging in Helmholtz Zentrum Munich, Oberschleißheim, Germany. His research interests are image reconstruction, deep learning, optoacoustic imaging, ultrasound imaging, biomedical instrumentation, and biomedical optics

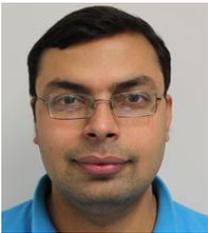
**Manish Arora:** Currently an associate professor at Department of Design and Manufacturing (DM), Indian Institute of Science (IISc). He obtained a Ph.D. in Applied Physics from the University of Twente, The Netherlands (2006). He has got 70+ patent and research publications both in national and international level. He was employed with University of Oxford, UK (2006-2010), GE Global Research (2010-2012) and Nanyang Technological University, Singapore (2012-2014). His research interests include biomedical devices, co-design, collaboration, open source in design, and quality manufacturing of medical devices. He is the principal investigator of UTSAAH Lab, which aims at developing affordable and accessible medical technology solutions for promoting universal healthcare. He teaches courses on Mechatronics, Design of Biomedical Devices and Systems at IISc.